**Versatile electronic phases enabled by intertwined multiple frustrations in antiferromagnetic two-dimensional semimetals**

Y. Fujisawa[1,2*], P. Wu[1,3*], T. Nakamura[1*], R. Okuma[1,4,5], T. Kato[1], B. R. M. Smith[1], D. Ueta[1,6], R. Kobayashi[7], N. Maekawa[7], C-H. Hsu[1], Chandan De[1], N. Tomoda[1], T. Higashihara[1,8], K. Morishita[1,7], K. Sumida[2], K. Miyamoto[2], T. Okuda[2], Z. Y. Wang[3], Y. Okada[1,5]

(* Equal contribution)

*[1] Quantum Materials Science Unit, Okinawa Institute of Science and Technology (OIST), Okinawa 904-0495, Japan.*

*[2] Research Institute for Synchrotron Radiation Science (HiSOR), Hiroshima University, Higashi-Hiroshima 739-0046, Japan.*

*[3] Department of Physics and Chinese Academy of Sciences Key Laboratory of Strongly-coupled Quantum Matter Physics, University of Science and Technology of China, Hefei, Anhui 230026, China.*

*[4] Institute for Solid State Physics (ISSP), The University of Tokyo, Kashiwa, Chiba 277-8581, Japan.*

*[5]JST, PRESTO, 4-1-8 Honcho, Kawaguchi, Saitama, 332-0012, Japan*

*[6] Institute of Materials Structure Science, High Energy Accelerator Research Organization, Tsukuba, Ibaraki 305-0801, Japan.*

*[7] Faculty of Science, University of the Ryukyus, Nishihara, Okinawa 903-0213, Japan.*

*[8] Department of Physics, Graduate School of Science, Osaka University, Toyonaka 560-0043, Japan.*

**Abstract**

Following the discovery of graphene, interest in van der Waals (vdW) materials has surged; however, advancing physics beyond graphene requires quantum vdW materials platforms that host versatile, strongly interacting many-body states. Here, using scanning tunneling microscopy and spectroscopy at 300 mK, we uncover multiple competing electronic states in the van der Waals metal $CeTe_3$: charge-ordered antiferromagnetic phases forming stripe and checkerboard orders. Remarkably, their competition is tuned by a modest in-plane magnetic field (~1.5 T), revealing strongly intertwined multiple frustrations involving antiferromagnetism, charge order, and Fermi-surface instabilities. Quasiparticle-interference imaging directly identifies the momentum-space origin of these competitions on the representative semimetal's Fermi surface. While the observations can be understood at a basic level in terms of Kondo coupling between localized Ce 4f moments and itinerant Te 5p electrons, our results reveal a much richer phenomenology: an unusually broad electronic reconstruction extending to an energy scale of roughly ± 30 meV from $E_F$, which realizes and deforms antiferromagnetic charge-ordered states and signals strongly correlated interactions beyond a weak-coupling description. Beyond establishing $CeTe_3$ as a model platform, our results demonstrate that competing instabilities in antiferromagnetic two-dimensional metals/semimetals generate versatile electronic-phases, opening a route to tunable nanoscale quantum states governed by the intertwined effects of correlation, symmetry, and topology.

## Introduction

### Broad scope

Fermi-surface (FS) reconstruction and pronounced reshaping of the density of states (DOS) near the Fermi level ($E_F$) are hallmarks of the emergence of distinct electronic phases, often associated with spontaneous symmetry breaking and resulting in exotic electronic functionality. While nesting of low-dimensional Fermi surfaces provides a conventional route to FS instabilities, the phenomenology becomes markedly richer when fermionic systems are subject to strong frustration, in which one state is selected from many competing possibilities. In contrast to quasi-one-dimensional systems—where nearly perfect nesting typically stabilizes an insulating state characterized by single-q spontaneous electronic ordering—quasi-two-dimensional systems with large FSs intrinsically exhibit imperfect nesting, giving rise to a wide variety of competing FS instabilities [1,2,3,4]. Furthermore, when frustrated Fermi-surface instabilities coexist with other frustration channels, such as spin and charge ordering, they can give rise to a variety of single-q and multi-q intertwined phases, ranging from fluctuating regimes to long-range ordered states, reminiscent of the rich phenomenology of strongly correlated systems such as high-temperature superconductors [5,6,7,8,9]. In particular, controlling magnetic phases characterized by multi-q, frustrated FS instabilities are of great interest for developing topological magnetism and associated exotic electronic transport arising from giant fictitious fields [10,11,12,13,14,15,16].

### Significance of developing AF vdW platform

In this context, such opportunities are particularly compelling in van der Waals (vdW) metals hosting frustrated antiferromagnetism (AF). Unlike q = 0 ferromagnets, AF states characterized by finite ordering wavevectors $q_m$ can, in general, exhibit greater versatility. Furthermore, their interplay with competing nesting and frustrated FS instabilities can yield unexpected intertwined phases that go beyond the phenomenology of ubiquitous charge order in vdW metals [17,18,19,20]. Such complex frustration and successive phase transitions, which typically drive the system toward reduced carrier density and more localized electronic states, can recruit additional many-body interactions, giving rise to exotic strongly correlated phases accompanied by electronic reconstructions over unusually broad energy scales around $E_F$, beyond weak-coupling descriptions [21]. However, despite these promising prospects and the ubiquitous observation of charge-density-wave (CDW) order in many non-magnetic vdW materials [22], how this interplay evolves across energy, momentum, and external control parameters remains poorly understood in antiferromagnetic vdW metals and semimetals.

### Rare earth tritelluride ($RTe_3$)

The rare-earth tri-telluride ($R$Te$_3$) family (**Fig. 1a**) has emerged as a versatile vdW quantum material platform [23]. These materials offer tunability through axial structural control [24], chemical substitution [17,25,26], exfoliation [27,28,29], external stimuli such as pressure, pulsed light, and RF field [30,31,32,33,34]. Also, interesting signatures of pressure-induced superconductivity [35,36,37], formation of Kramers nodal line band structures [38], and possible links to Higgs mode analogous in high-energy physics [39,40] are shown. Across the family, the initial two bands formed from Te 5p orbitals on square-net layers are folded by the potential of the block layers, resulting in a relatively simple parent FS in the absence of any CDWs (**Fig. 1c**). On top of this, CDW1—which is common across all $RTe_3$ compounds (**Fig. 1b**)—introduces additional band folding and associated frustrating quasi-nesting (**Fig. 1d**). Intriguingly, Landau-level spectroscopy reveals well-defined electron–hole pockets emerging from this otherwise complex folded structure (see the yellow shaded area in **Fig. 1d**) [41]. Thus, the significance of considering the instability of the Fermi surface in such representative

semimetals—characterized by coexisting small electron and hole pockets—becomes effective. This simple viewpoint captures the essence of the parent electronic state that incorporates the CDW1 characteristics common to the $RTe_3$ family and provides a crucial foundation for deepening our understanding of emergent low-temperature phases.

### Significant position of $CeTe_3$

While early studies predominantly focused on high-temperature charge-ordered states (CDW1 and CDW2) [42], recent efforts have highlighted intertwined orders that emerge within antiferromagnetic phases below the onset of long-range antiferromagnetic order, i.e., the Néel temperature ($T_N$) [43,44,45,46]. In this context, $CeTe_3$ stands out as a singular member that enables a controlled entry point into intertwined charge–spin physics in two dimensions [17,47,48,49,50]. In $CeTe_3$, a $4f^1$ moment develops long-range AF order at $T_N \approx 1.5$ K and undergoes a spin-flop transition at $B_{flop} \approx 1.5$ T by in-plane field, which reorients the in-plane Néel vector (**Fig. 1e,f**). Concomitantly, $CeTe_3$ exhibits a pronounced and unusually broad spectral reconstruction across $T_N$, extending the energy scale of approximately ± 30 meV from $E_F$ (shaded region in **Fig. 1g,h**), a feature absent in other $RTe_3$ such as $TbTe_3$. The fact that this magnetic transition manifests directly as a modification of the electronic states near $E_F$ indicates an intimate coupling between magnetism and the low-energy electronic structure, likely arising from moderate Kondo coupling between the Ce $4f^1$ moments and Te 5p electrons [47]. This interpretation is consistent with previous specific-heat measurements, which indicate a moderately screened Ce moment relative to 1 $\mu_B$ [17]. Further insight is provided by neutron-scattering experiments, which reveal a magnetic propagation vector $q_m$ = (h, l) = (1/6, ±1/3) at B = 0 [50]. This information provides a valuable reference for comparison with the charge-propagation vector $q_c$, if determined, thereby enabling a deeper understanding of the underlying spin–charge coupling mechanism. Despite the promising opportunities offered by $CeTe_3$, the microscopic origin of these effects—and how charge and magnetic orders are selected and coupled—remains unresolved, motivating local and momentum-resolved spectroscopic studies under controlled temperature and magnetic-field conditions.

### Position of this study

Here, using spectroscopic scanning tunneling microscopy (STM), we track the evolution of the electronic structure of $CeTe_3$ across $T_N$ and $B_{flop}$ (**Fig. 1e,f**) [17]. On the established parent state hosting only CDW1 for $T > T_N$ (**Fig. 1ij**) [51,52,53], we identify three additional charge modulations with wavevectors (0.33, 0), (0.19, ±0.19), and (0, 0.08) that emerge within the AF state. All measurements employed a non-magnetic STM tip, enabling direct access to intrinsic DOS reconstructions across $T_N$ and $B_{flop}$ without convolution from spin-dependent tunneling matrix elements inherent to spin-polarized STM. By combining real-space STM with quasiparticle-interference (QPI) imaging, focusing on magnetically ordered states below $T_N$, we demonstrate microscopic insights into versatile antiferromagnetic charge order and its competition in momentum space in $CeTe_3$. Based on these insights, we discuss and provide a foundation for future efforts to control and exploit intertwined orders—and their field-tunable competition—applicable broadly to AF vdW materials.

## Results and Discussions

### Emergence of magnetically driven charge order across $T_N$

We first identify magnetically driven charge orders from STM measurements performed across the AF transition at $T_N \approx 1.5$ K under zero magnetic field (see **Methods** for **sample characterization**). At 4 K, the topographic image (**Fig. 2a**) exhibits a unidirectional modulation with a Fourier wavevector $q_{CDW1}$ = (0, 0.28) (**Fig. 2b**), consistent with the well-established CDW1 phase of $CeTe_3$ (**Fig. 1b**). Upon cooling to 300 mK, qualitatively distinct stripe-like modulations emerge. This new modulation is oriented perpendicular to CDW1 and is characterized by charge-propagation vectors (0.33, 0) and (0, 0.08) (**Figs. 2c,d**). The same periodicities are clearly resolved in energy-dependent dI/dV maps in real space (**Fig. 2e**) and are consistently captured in Fourier space (**Figs. 2f–i**), visualizing their non-dispersive nature with energy. Because the wavevector (0.33, 0) resembles the conventional CDW2 reported in other $RTe_3$ compounds, establishing whether this modulation is intrinsically linked to antiferromagnetism requires a detailed temperature-dependent study.

For efficient tracking of the relevant charge orders, we note that while the modulation at (0.33, 0) can be visualized over a relatively wide bias window of ±50 mV, topographic images acquired within a narrower bias window of ±10 mV are optimal for resolving the modulation with wavevector (0, 0.08) (see **Methods** and **Supplementary Note 2**). Accordingly, to track the simultaneous evolution of both charge orders, we focus on topographic images acquired, for example, at −10 mV. Notably, by following this strategy, while CDW1 shows little change across $T_N$, the additional charge modulations characterized by wavevectors (0.33, 0) and (0, 0.08) collectively exhibit a clear onset below $T_N$ (**Fig. 2j**). Combined with their suppression above the spin-flop field $B_{flop}$ (discussed later), the modulation (0.33, 0) is established as a charge order intrinsically intertwined with antiferromagnetism, which we denote $CDW2_{mag}$. Similarly, the modulation at (0, 0.08), which also develops exclusively below $T_N$, is denoted $CDW3_{mag}$ (**Fig. 2d,j**). Throughout this work, the subscript “mag” is used to indicate charge orders that emerge only within magnetically ordered states.

### Field-induced emergence of $CBC_{mag}$ above $B_{flop}$

Applying an in-plane magnetic field induces a pronounced reorganization of charge order within the AF phase. Similar to the previous study [17], in-plane field is applied along horizontal c-axis, which is perpendicular to the CDW1 direction (**Fig. 1f**). At B = 2.0 T, exceeding the spin-flop field $B_{flop} \approx 1.5$ T, the topographic image and its Fourier transform (**Figs. 3a,b**) reveal the emergence of two orthogonal charge modulations with wavevectors (0.19, ±0.19) (red and yellow circles in **Fig. 3b**). These modulations are directly visualized in real-space dI/dV maps (**Fig. 3c**) and corroborated by Fourier line profiles (**Figs. 3d-g**). The absence of measurable energy dispersion indicates that these modulations originate from static ordering (**Fig. 3d,e**). Because this charge order appears only within the magnetically ordered phase and is stabilized by an applied magnetic field, we refer to this state as $CBC_{mag}$.

It should be noted that the charge-propagation vectors associated with $CDW2_{mag}$, $CDW3_{mag}$, and $CBC_{mag}$ connect nearly nested regions of the Fermi surface (**Fig. 1j**). As a result, competition among these orders is naturally expected: stabilization of one charge order partially gaps the Fermi surface, thereby reducing the fermionic energy gain available to competing instabilities. Here, we use the term “nesting” in a broad sense, referring not only to ideal parallel Fermi-surface segments but also more generally to wavevectors connecting regions with a relatively large joint density of states. Within the scope of this study, deviations from ideal nesting conditions are not emphasized, provided that partial gap opening still yields a substantial net fermionic energy gain.

## Field-tuned competition of antiferromagnetic charge orders

The relevant competition is directly visualized through the magnetic-field dependence of the Fourier intensities associated with $CDW2_{mag}$, $CDW3_{mag}$, and $CBC_{mag}$ (**Fig. 3h**). While $CDW3_{mag}$ is optimally visualized using bias voltages of approximately ±10 meV (**Figs. 2c,d**), we find that the competition between $CDW2_{mag}$ and $CBC_{mag}$ is more efficiently captured in topographic images acquired at larger bias voltages on the order of ±50 meV (see **Methods** and **Supplementary Note 2**). As shown in **Fig. 3h**, the $CDW3_{mag}$-related signal is nearly buried within the background noise at a bias voltage of +50 meV at B = 0 (see dashed line). With increasing magnetic field, however, the enhancement of the $CDW3_{mag}$ signal enables the simultaneous visualization of $CDW2_{mag}$, $CDW3_{mag}$, and $CBC_{mag}$ within a single topographic dataset, allowing a consistent comparison across the entire field range. Here, to eliminate extrinsic anisotropic contrast arising from the STM tip, the Fourier intensities are normalized by the corresponding Bragg peaks: (0,1) for CDW1 and $CDW3_{mag}$, (1,0) for $CDW2_{mag}$, and (1,±1) for $CBC_{mag}$. Although $CDW3_{mag}$ evolves relatively smoothly with the magnetic field, a weak discontinuity is nevertheless observed across the spin-flop field $B_{flop}$. More strikingly, as the magnetic field increases from zero, the intensity of $CDW2_{mag}$ is sharply suppressed, while $CBC_{mag}$ is simultaneously enhanced across $B_{flop}$, giving rise to a pronounced anti-correlation between the two orders.

## Density-of-states reconstruction and energy hierarchy across $B_{flop}$

The energy scale associated with the competition among multiple magnetic charge-ordered states is directly reflected in the magnetic-field dependence of the tunneling spectra dI/dV(E) (**Fig. 4a**). Using the bimodal color scheme (blue and red) shown in **Fig. 4b**, the spectra can be clearly classified into two distinct groups below and above the spin-flop field $B_{flop}$. To quantify the characteristic energy scale, we directly compare representative spectra measured at B = 0 T (< $B_{flop}$) and B = 2 T (> $B_{flop}$). Taking the derivatives of the dI/dV spectra and overlaying them (**Fig. 4c**) highlights the energy range over which the density of states (DOS) is most strongly modified. The gray curve in **Fig. 4c** represents a reference spectrum, denoted $d^2I/dV^2_{B\text{-ave}}$, obtained by averaging all spectra shown in **Figs. 4a** and **4b**. This comparison demonstrates that the dominant spectral changes are confined to an energy window of approximately ±20 meV around $E_F$. Consistently, the field evolution, plotted as $d^2I/dV^2(B) - d^2I/dV^2_{B\text{-ave}}$, further reveals a pronounced DOS reconstruction across $B_{flop}$ within the same energy range (**Fig. 4d**).

Rather than emphasizing fine spectral structures visible in **Fig. 4d**, the most robust outcome of the magnetic-field-dependent analysis is the identification of this characteristic energy window. In two-dimensional multiband systems, superlattice formation driven by Fermi-surface nesting typically produces complicated fine spectral features within the reconstructed energy range. Moreover, in $LaTe_3$—where only CDW1 order exists (**Fig. 1b**)—electron–boson coupling generates dip-like structures near $E_F$ that are not directly related to the CDW condensation energy [41]. Similar spectral complexity is therefore expected in $CeTe_3$. In this context, the energy scale of the DOS modulation across $B_{flop}$ provides a more reliable indicator than the detailed fine structure within the soft gap.

Combining the field-dependent evolution of the raw spectra (**Fig. 4b**) with the derivative analysis yields a consistent picture: the electronic structure changes across $B_{flop}$ on an energy scale of approximately ±20 meV. Notably, the dominant spectral modifications extend across $E_F$, indicating that the DOS reconstruction reflects a fermionic energy gain or loss associated with the competing phases. As schematically illustrated in **Fig. 4e**, the strongest manifestation of this energy competition appears near −10 meV, highlighted by the shaded regions in **Figs. 4b** and **4c**.

**Quasiparticle interference pattern**

To deepen our understanding of the electronic competition in momentum space, we investigate quasiparticle interference (QPI). Motivated by the DOS reconstruction observed at 300 mK around −10 meV across $B_{flop}$ (**Fig. 4e**), we concentrate on comparing high-resolution QPI data taken at 0 and 2 T at the same temperature and energy (**Figs. 5a,b**). To isolate the field-induced evolution across $B_{flop}$, the QPI pattern at 0 T is subtracted from that at 2 T, yielding a differential QPI map (**Fig. 5c**), which reveals a characteristic suppression (blue) and enhancement (red) of QPI intensity upon increasing the magnetic field. To interpret this red/blue contrast, we consider nesting vectors associated with $CDW2_{mag}$, $CDW3_{mag}$, and $CBC_{mag}$ on approximate constant-energy contours constructed from ARPES measurements of the parent band structure above $T_N$ (**Fig. 1k**). At B = 0 T, nesting associated with $CDW2_{mag}$ (blue arrows in **Fig. 5d**) leads to partial gap opening predominantly on the inner pocket (**Fig. 5e**), whereas at B = 2 T, nesting associated with $CBC_{mag}$ and $CDW3_{mag}$ collectively produces pronounced gap opening on the outer pocket (**Figs. 5f,g**).

The essential physics underlying the experimental QPI difference between 0 and 2 T is most clearly manifested near q = (0,0). In this region, the QPI is governed by scattering between quasi-one-dimensional, predominantly 5p-orbital-derived Fermi surfaces (**Fig. 1c,d**), whose parallel constant-energy contours generate features that cross q = (0,0). Within this framework, the dominant scattering shifts from the outer pocket at B = 0 T to a shorter vector within the inner pocket at B = 2 T, as indicated by the solid and dashed arrows in **Figs. 5h** and **5i**. This shift naturally results in a characteristic change in the momentum-space extent of the QPI across q = (0,0) (**Fig. 5j**), which is directly captured in the experimental differential QPI map (**Fig. 5c**).

The validity of this picture is further supported by its ability to account for the QPI modulation associated with the development of $CDW3_{mag}$ at higher magnetic fields (**Fig. 3h**). In our model, the corresponding QPI signal is expected to deform and become suppressed as $CDW3_{mag}$ is stabilized, consistent with the k-space reconstruction shown in the magnified regions of **Figs. 5h** and **5i**. The propagation vector q = (0, 0.08) associated with $CDW3_{mag}$ is indicated by arrows in **Figs. 5a–c**, around which extended QPI features are observed. While a prominently extended QPI signal is present at B = 0 T, this signal is strongly suppressed at B = 2 T, concomitant with the enhanced $CDW3_{mag}$ intensity localized at q = (0, 0.08) (highlighted regions in **Figs. 5a,b**). Accordingly, the theoretically expected disappearance (blue-coded) of an arc-like extended QPI feature around q = (0, 0.08) at 2 T (**Fig. 5j**) is faithfully reproduced in the experimental differential QPI map (**Fig. 5c**). Taken together, the QPI deformations observed between B = 0 and 2 T—particularly near q = (0,0)—are consistently explained within a minimal modeling framework incorporating $CDW2_{mag}$, $CDW3_{mag}$, and $CBC_{mag}$, providing direct momentum-space evidence for the competition and coexistence of these charge-ordered states.

**Magnetic propagation vectors and spin–charge intertwining**

We next discuss the relationship between the charge modulation vectors $q_c$ and the magnetic propagation vectors $q_m$. A natural microscopic link between these degrees of freedom is the Kondo exchange coupling between itinerant Te 5p electrons and localized Ce 4f moments, as suggested by previous experimental and theoretical studies [28,29,47,54,55,56]. At zero magnetic field below $T_N$, neutron-scattering experiments have revealed antiferromagnetic order characterized by two propagation vectors, $q_{m1} \approx (1/6, +1/3)$ and $q_{m2} \approx (1/6, -1/3)$ [50]. Theoretical models incorporating finite Kondo exchange predict that charge modulations can emerge as linear combinations of such magnetic propagation vectors [12,15,16]. This naturally yields two candidate charge wavevectors: $q_{m1} + q_{m2}$ = (1/3, 0) and $q_{m1} - q_{m2}$ = (0, 2/3). Among these, the latter vector does not satisfy the nesting condition and therefore does not lead to an observable magnetic charge modulation. In contrast, the former vector efficiently connects nested regions of the Fermi surface, allowing a fermionic energy gain that stabilizes a magnetic charge-ordered state, $CDW2_{mag}$. Consequently, at zero magnetic field, $CDW2_{mag}$ emerges as an intertwined spin–charge order that selects a single-q charge propagation vector, despite being supported by an underlying double-q magnetic structure at (1/6, ±1/3).

Under finite magnetic fields, neutron-scattering information remains limited, and direct determination of the magnetic structure is an important subject for future work. Nevertheless, the emergence of a double-q charge order with wavevectors (±0.19, 0.19) (**Fig. 3h, right**) places strong constraints on the underlying magnetic configuration. Assuming that Kondo exchange remains operative, the realization of double-q charge order implies the stabilization of a distinct multi-q magnetic state that enables an electronic energy gain at these wavevectors. One plausible scenario involves double-q antiferromagnetic propagation vectors $q_{m1}$ = (0.19, 0) and $q_{m2}$ = (0, 0.19), which yield $q_{m1} + q_{m2}$ = (0.19, 0.19) and $q_{m1} - q_{m2}$ = (−0.19, 0.19). Alternatively, when a finite net magnetic moment develops under a magnetic field above $B_{flop}$, the emergence of a $q_m = 0$ ferromagnetic component would allow the magnetic and charge propagation vectors to coincide ($q_c = q_m$), naturally stabilizing the observed (±0.19, 0.19) charge order. In either case, $CBC_{mag}$ represents a charge order that is intrinsically intertwined with a distinct multi-q magnetic structure. Although the precise spin texture of this field-induced phase remains unresolved directly from neutron diffraction measurements, recent theoretical studies of easy-plane magnets predict a wide variety of stable multi-$q_m$ states under magnetic fields, including vortex-crystal phases [56].

By contrast, the wavevector of $CDW3_{mag}$ at (0, 0.08), which weakly exists even at zero magnetic field (**Fig. 2d**), cannot be straightforwardly accounted for by any magnetic Bragg peaks reported in existing neutron-scattering studies. Unlike $CDW2_{mag}$ and $CBC_{mag}$, we therefore cannot unambiguously identify a mechanism in which this charge order is directly coupled to a known magnetic order. Nevertheless, the possibility that a corresponding magnetic peak has remained undetected due to limited experimental sensitivity or measurement conditions cannot be excluded, motivating future high-resolution neutron-scattering investigations.

**Underlying microscopic interactions**

Although Kondo exchange plays an essential role in mediating the coupling between localized spins and itinerant electrons, our observations point to substantially richer physics beyond this simple picture. The reported Kondo temperature is on the order of 10 K [47], whereas the magnetic ordering temperature and the spin-flop field observed in this study are only about 1.5 K and 1.5 T, respectively [17]. These energy scales are therefore far too small to account for the pronounced reconstruction of the density of states observed here, which extends over approximately ±30 meV from $E_F$ across $T_N$ (**Fig. 1g,h**) and a comparable ±20 meV across $B_{flop}$ (**Fig. 4d**). This clear separation of energy scales indicates that Kondo exchange alone cannot be the dominant driving interaction. Instead, additional interactions must play a decisive role in shaping the low-energy electronic structure, which should be considered an intriguing feature of $CeTe_3$. Notably, the energy scale of the band deformation identified here (**Fig. 4**) is comparable to the dispersion kink associated presumably with electron–phonon coupling revealed by Landau-level spectroscopy in the nonmagnetic single-CDW system $LaTe_3$ [41]. This comparison demonstrates that similarly high-energy electronic coupling is already intrinsic to the Te-based layered electronic structure, even in the absence of 4f moments. In this context, Coulomb interactions between the small electron and hole pockets in this semimetallic system—an electronic configuration closely analogous to that of $CeTe_3$ (**Fig. 1d** and **Fig. 5f)**—are also a candidate driving source of large energy-scale Fermi-surface deformations reminiscent of an excitonic band instability [20,57,58]. Although the present system is not insulating, such effects are expected to be strongly enhanced in van der Waals materials, where reduced dimensionality and weakened electronic screening amplify interaction effects. Taken together, these considerations establish $CeTe_3$ as an exceptional two-dimensional platform in which multiple interactions—Kondo exchange, Coulomb correlations, and electron–phonon coupling—cooperatively generate versatile antiferromagnetic charge orders. Through this interplay, tunable and strongly correlated electronic states emerge on a semimetallic Fermi surface, extending well beyond a simple Kondo-driven picture.

**Outlook and Future Perspectives**

From a real-space perspective, this rich landscape offers fertile ground for exploring exotic multi-$q_m$ magnetic orders. Such textures are widely regarded as a key ingredient for realizing topologically nontrivial magnetic states and can give rise to large emergent electromagnetic responses, including enhanced transport signals associated with fictitious magnetic fields [10]. Beyond multi-$q_m$ magnetism, the coexistence of multiple charge orders is also expected to induce unconventional symmetry breaking; for example, nontrivial relative phase shifts between charge orders may generate exotic in-plane inversion symmetry breaking [59]. These possibilities naturally motivate future studies combining microscopic probes with Hall and non-reciprocal transport measurements.

From a complementary momentum-space viewpoint, the realization of versatile multi-q antiferromagnetic charge order on a semimetallic Fermi surface opens new opportunities to explore unconventional electronic structures. In this regime, competing Fermi-surface instabilities may drive the emergence of Dirac-like dispersions, band inversion, and rich momentum-space textures of electronic wave functions. Moreover, the strong spin–charge intertwining uncovered in $CeTe_3$ provides a new pathway toward unconventional quantum criticality. Whereas in related $RTe_3$ compounds such as $TbTe_3$ and $DyTe_3$ superconductivity emerges near the CDW2 critical point, where magnetism and charge order are largely decoupled [23,35,36], the magnetically intertwined charge orders identified here point to a qualitatively distinct critical regime and suggest new routes toward unconventional superconductivity driven by coupled spin and charge degrees of freedom.

Finally, guided by the general concept of moiré modulation—where the superposition of periodic structures generates new wave vectors through linear combinations—our observation of multiple electronic propagation vectors can be viewed, at a conceptual level, as a manifestation of versatile charge–spin intertwined electronic moire states, rather than a superposition tied to crystalline lattice vectors [60,61]

**Summary**

In summary, our results establish $CeTe_3$ as a versatile platform for realizing multi-q antiferromagnetic charge orders, accompanied by large-energy-scale electronic reconstructions characteristic of strongly correlated systems. The observed phenomenology originates from the intertwining of multiple frustration channels—antiferromagnetism, charge ordering, and competing Fermi-surface instabilities—within a semimetallic state. From a conceptual perspective, this study demonstrates that van der Waals antiferromagnetic semimetals offer promising opportunities to explore previously inaccessible, tunable quantum phases in which topology and strong electronic correlations are intrinsically intertwined in two dimensions. From a functional viewpoint, these results provide guiding principles for designing quantum materials that integrate multiple input and output channels, enabling disproportionately large electronic responses to modest external stimuli at the nanoscale in versatile electronic moire phases.

## Methods

### Sample growth and characterization

Single crystals were grown following procedures similar to those described in Ref. [17]. Phase purity and stoichiometry are well characterized by XRD and EDS. In the previous report, two characteristic temperatures ($T_{N1}$ and $T_{N2}$) were pointed out [17]. $T_{N1}$ corresponds to short-range magnetic fluctuation, and $T_{N2}$ corresponds to long-range antiferromagnetic order. Note that the $T_N$ in this report corresponds to $T_{N2}$ in the literature. We also note that the slightly higher $B_{flop}$ inferred from STM/STS at 300 mK compared with magnetometry at 0.5 K **(Fig. 1i**) is naturally attributed to the temperature difference rather than field misalignment (see calibration of B-field for **Supplementary Note 5**).

### STM/STS

General experimental conditions

Samples were cleaved in situ under ultra-high vacuum (UHV) conditions at room temperature and transferred directly to the STM head without exposure to air. Based on our experience, cleaving at either 77 K or room temperature does not affect the main conclusions of this study. All samples used in this study were cleaved in situ under UHV at room temperature. Chemically etched W tips were annealed at high temperature in UHV to remove surface oxides. Within the relevant bias voltage range, tip sharpness and stability, as well as density-of-states (DOS) flatness, were carefully calibrated on Au(111) prior to sample measurements. The non-magnetic nature of the STM tip was verified through dedicated control experiments, ensuring that the observed magnetic-field-dependent dI/dV signals originate from intrinsic changes in the sample DOS rather than from spin-dependent tunneling matrix elements. No evidence of tip-condition changes was observed throughout the measurements presented here. Differential conductance spectra were acquired using a standard lock-in technique with a modulation frequency of 961 Hz.

Detection of charge order from topography and set-point effect

Acquiring fully energy-resolved STS datasets over the entire magnetic-field range is extremely time-consuming and significantly increases the risk of changes in the tip or surface condition. To ensure reliable data acquisition within a realistic experimental timescale, we therefore use STM topographic imaging as the primary method to track $CDW2_{mag}$, $CDW3_{mag}$, and $CBC_{mag}$. While $CDW3_{mag}$ is best visualized with bias voltages of approximately ±10 meV, the competition between $CDW2_{mag}$ and $CBC_{mag}$ is more clearly captured at ±50 meV. As shown in **Fig. 3h**, at B = 0 the $CDW3_{mag}$-related signal is nearly buried in the background noise at a bias voltage of +50 meV (see dashed line). With an increasing magnetic field, however, the $CDW3_{mag}$ signal enhancement allows simultaneous visualization of $CDW2_{mag}$, $CDW3_{mag}$, and $CBC_{mag}$ in a single topographic image. Additional experimental details are provided in the **Supplementary Note 2**.

dI/dV spectra

To avoid misleading data interpretation, each dI/dV spectrum is spatially averaged over regions over at least 30x30 $nm^2$. This averaging procedure also suppresses quasiparticle-interference modulations. Thermal- and magnetic-field-induced extrinsic drift of the probed surface area is carefully corrected, ensuring that spectra acquired under different conditions correspond to essentially the same spatial location.

Strictly speaking, normalization of dI/dV spectra involves inherent uncertainty, which limits a fully quantitative analysis of absolute DOS deformation (**Fig. 4**). Nevertheless, because all spectra are acquired under identical set-point conditions and instrumental energy resolution, they reliably capture relative changes in the DOS near the Fermi level. When comparing spectra acquired at different temperatures, thermal broadening must also be accounted for. To address this qualitatively, we simulate a Gaussian-broadened spectrum corresponding to 4 K based on the 0.3 K, B = 0 data. This comparison indicates that thermal broadening plays only a minor role in shaping the main conclusions of this study. Under our experimental conditions, the electronic temperature is sufficiently low to be negligible compared with the sample temperature. Further experimental details are provided in the **Supplementary Note 3**.

Quasiparticle interference (QPI) pattern measurement

Within the scope of this study, we focus on the antiferromagnetic state below $T_N$ at 300 mK. A systematic temperature-dependent investigation is deferred to future work, where it is expected to be particularly informative. Notably, in the temperature regime above $T_N$, presumably below ~10 K [47], we observe indications of an unusual matrix-element effect associated with fluctuating orders [53]. Clarifying its origin requires substantial additional experimental efforts and is therefore beyond the scope of the present study. Further details are provided in **Supplementary Note 6**.

Measurement stability, reproducibility

Measurement stability and reproducibility are verified throughout this study. In particular, before and after acquiring the complete set of magnetic-field- and/or temperature-dependent datasets (**Figs. 2–5**), we carefully confirm the reproducibility of both tunneling spectra and topographic images, with no discernible changes observed. This demonstrates that the reported results are robust against instrumental drift and measurement history. In addition, the robustness of the main conclusions is further confirmed using more than ten independently prepared STM tips on different sample surfaces (see also **Supplementary Note 4**).

## Acknowledgement

We thank Hayami and Niimi for the fruitful discussion.

We thank the University of the Ryukyus Research Facility Center for the use of the helium liquefier.

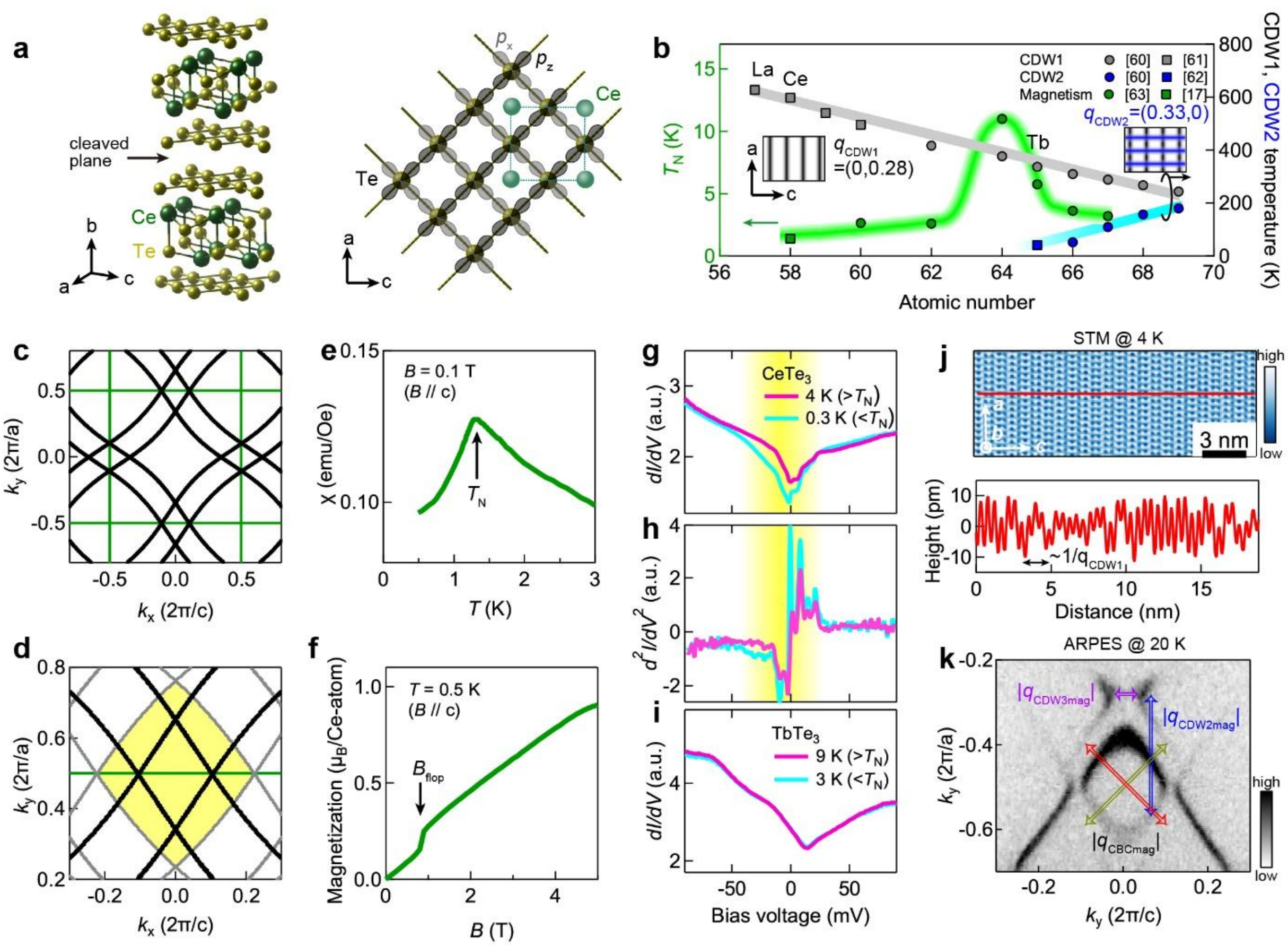

**Fig. 1. Unique position of $CeTe_3$ among the $R$$Te_3$ family**

(a) Crystal structure of $CeTe_3$ and top view of the lattice. The arrow in the left panel indicates the cleaved plane.

(b) Schematic phase diagram of the $RTe_3$ family, illustrating two well-established charge density wave (CDW) transitions (CDW1 and CDW2) and the Néel temperature ($T_N$), based on Ref. [17,62,63,64,65].

(c,d) Simplified FS topology derived from a tight-binding model considering Te $5p_x$ and $5p_Y$ orbitals forming a square net and block layer band folding (c), and additional folding due to the CDW1 formation (d). Green lines represent the Brillouin zone for the unit cell.

(e,f) Temperature (e) and magnetic field (f) dependence of the magnetization in $CeTe_3$. Black arrows indicate magnetic transition points. The data is adopted from our previous study [17].

(g,h,i) Comparison of the spatially averaged *dI/dV* spectra across $T_N$ for $CeTe_3$ (g), its derivative (h) and the spectra for $TbTe_3$ (i), respectively. The set-point is 90 mV/2.5 nA with lock-in amplitude of 0.4 mV (961 Hz) for $CeTe_3$ and -50 mV/400 pA with lock-in amplitude of 1 mV (961 Hz) for $TbTe_3$.

(j) The topographic image of $CeTe_3$ at 4K, representing CDW1 and a line profile along the red line.

(k) Constant energy contour at the Fermi level obtained by ARPES at 20 K.

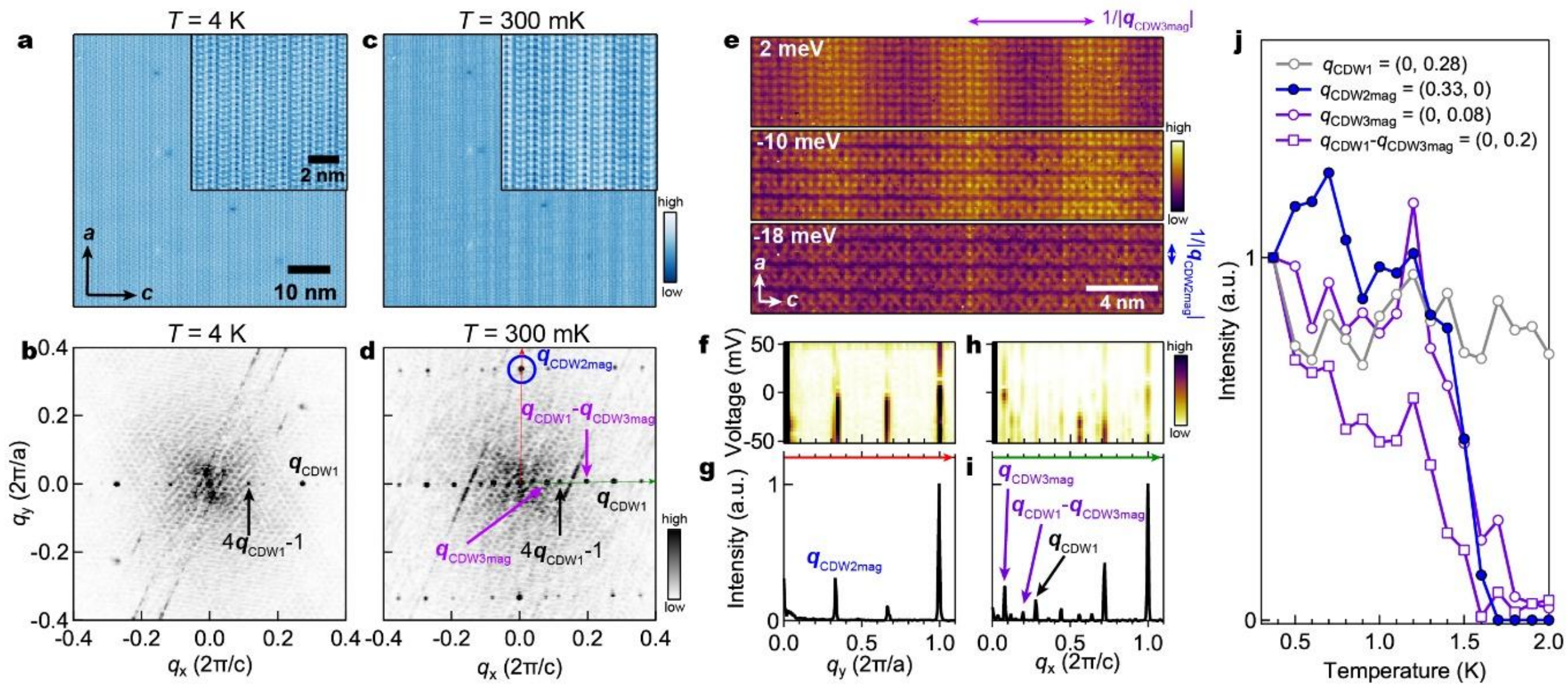

**Fig. 2. Temperature dependence of CDW2mag**

(a, b) The topographic image at 4 K (a) and its corresponding Fourier transform (FT) image (b).

(c, d) The topographic image at 300 mK (c), and its FT image (d). The setpoint conditions for (a) and (c) are similar (50 mV/4 nA and 50 mV/5 nA, respectively). All peaks corresponding in (b) and (d) are identified as a linear combination of relevant peaks of charge orders and crystal structures. See **Supplementary Note 1** for the full assignment of the FFT components.

(e) The *dI/dV* maps acquired over a 10 × 10 nm² area at various energies from -50 mV to 0 mV (setpoint conditions: 50 mV/5 nA with a lock-in modulation of 2 mV). The blue and purple arrows indicate the newly appeared periodicity corresponding to $q_{CDW2mag}$ and $q_{CDW3mag}$, respectively.

(f,h) The energy evolution of the line-cut profiles along $q_y$ and $q_x$ directions obtained from the dI/dV maps (**Fig. 2e**) (see the arrow indicated in d).

(g,i) Line-cut profiles of the Fourier signal along the $q_y$ and $q_x$ direction at 300 mK. The signal is obtained from the topographic image (**Fig. 2d**).

(j) Temperature dependence of the Fourier components at (1/3, 0), (0, 0.28), and (0, 0.2) extracted from the topographic images. The setpoint is -10 mV/500 pA.

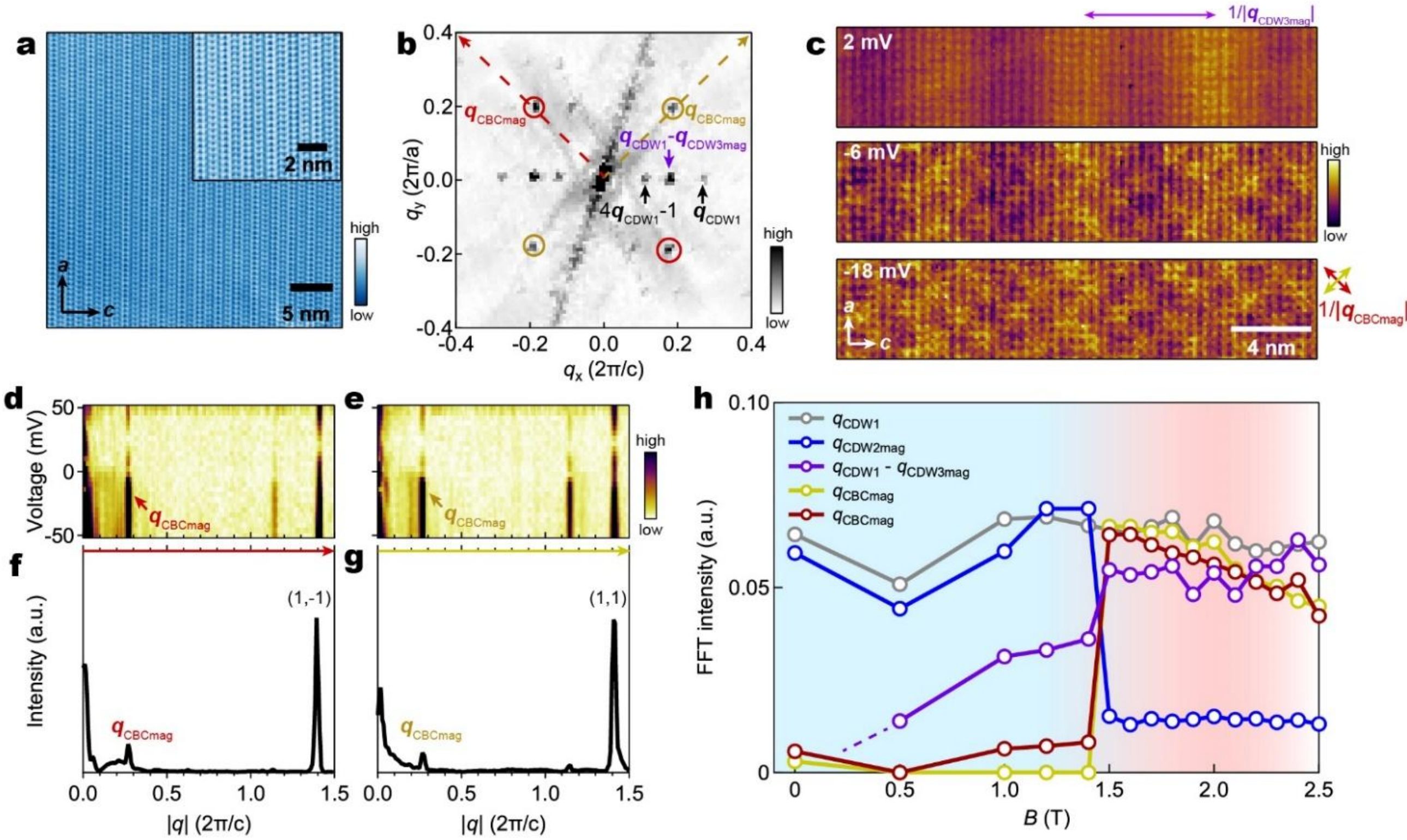

**Fig. 3. Competing CDW2$_{mag}$ and CBC$_{mag}$ at 300 mK across the spin-flop transition at B$_{flop}$ under an in-plane magnetic field applied along the c-axis.**

(a) STM topograph acquired at $B$ = 2 T applied along the $c$-axis at 300 mK.

(b) FT image corresponding to (a). The measurement conditions for (a) are 50 mV/2.5 nA. See **Supplementary Note 1** for the full assignment of the FFT components.

(c) The *dI/dV* mapping at various energies as indicated inside. The set point is 50 mV/2.5 nA with the lock-in modulation of 2 mV.

(d,e) The energy evolution of the line-cut profiles along (1,-1) and (1,1) directions, respectively (see the red and yellow arrows indicated in b).

(f,g) Line-cut profiles of the Fourier signal of the STM topograph (b) along (1,-1) and (1,1) directions, respectively.

(h) Magnetic field evolution of the FT peak intensities at $q_{CDW1}$, $q_{CDW2mag}$, $q_{CDW1}$-$q_{CDW3mag}$, and $q_{CBCmag}$, which are normalized by the Bragg peaks at ($q_y$,$q_x$) = (0,1), (1,0), (0,1), (±1,1), respectively, extracted from topographic images taken with the set-point of +50 mV/2.5 nA. This choice of normalization factors for each charge modulation is not to compare the amplitude between them, but to reduce the effect of a small tip anisotropy.

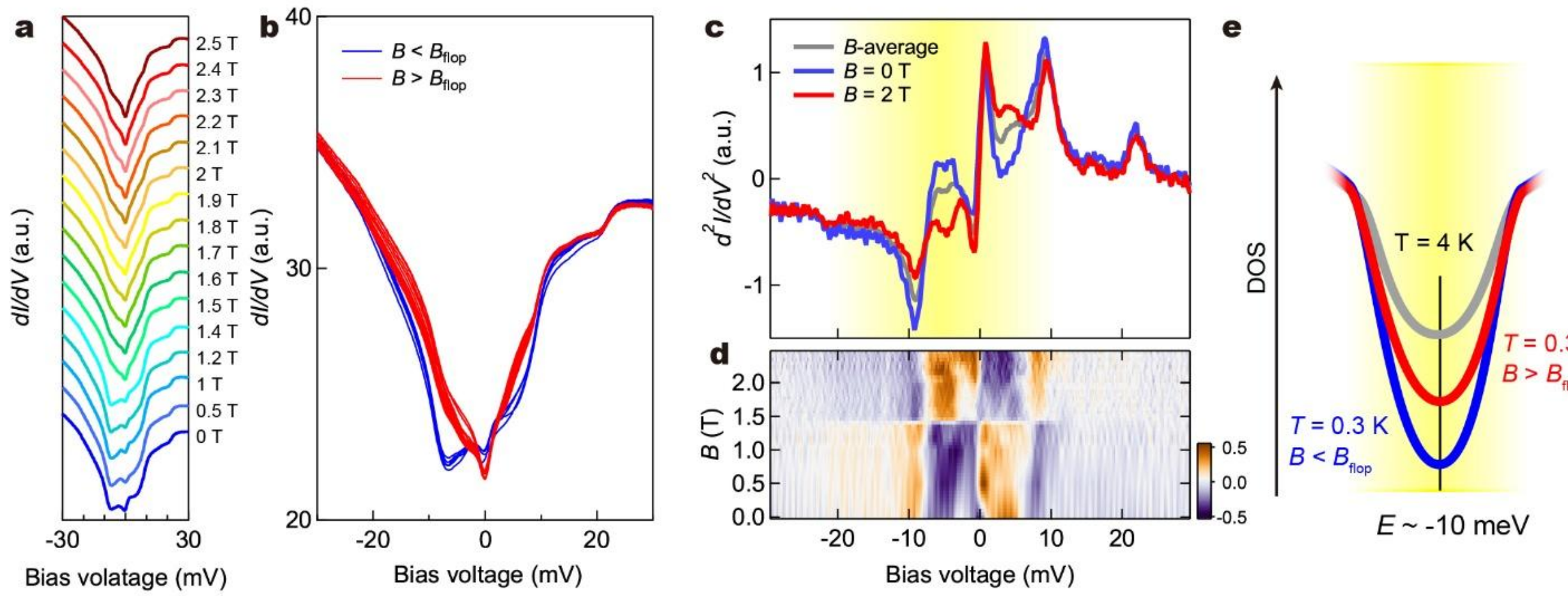

**Fig. 4. *dI/dV* deformation and competing nesting conditions at 300 mK under an in-plane magnetic field applied along the c-axis.**

(a) Evolution of the spatially averaged *dI/dV* spectra under in-plane magnetic fields at 300 mK. Spectra are vertically offset for clarity. The measurement condition is 50 mV/2.5 nA with the lock-in modulation of 0.8 mV.

(b) Representative spectra above and below the spin-flop field ($B > B_{flop}$ and $B < B_{flop}$), highlighted in two distinct colors.

(c) Derivative of the dI/dV spectra ($d^2I/dV^2$) at B = 0 and 2 T. The yellow-highlighted region indicates the energy window where substantial spectral renormalization is observed. The gray line indicates the magnetic field averaged spectrum ($d^2I/dV^2_{B\text{-ave}}$, obtained by averaging six curves (0, 0.5 ,1 , 1.5, 2, 2.5 T), where three are representative for $B < B_{flop}$ (0, 0.5, 1 T) and the other three are representative for $B > B_{flop}$ (1.5, 2.0, 2.5T).

(d) The Magnetic field dependence of $d^2I/dV^2(B)$-$d^2I/dV^2_{B\text{-ave}}$, revealing the pronounced spectral evolution within -20 meV to +10 meV.

(e) Schematic spectral evolution with changing temperature and magnetic field.

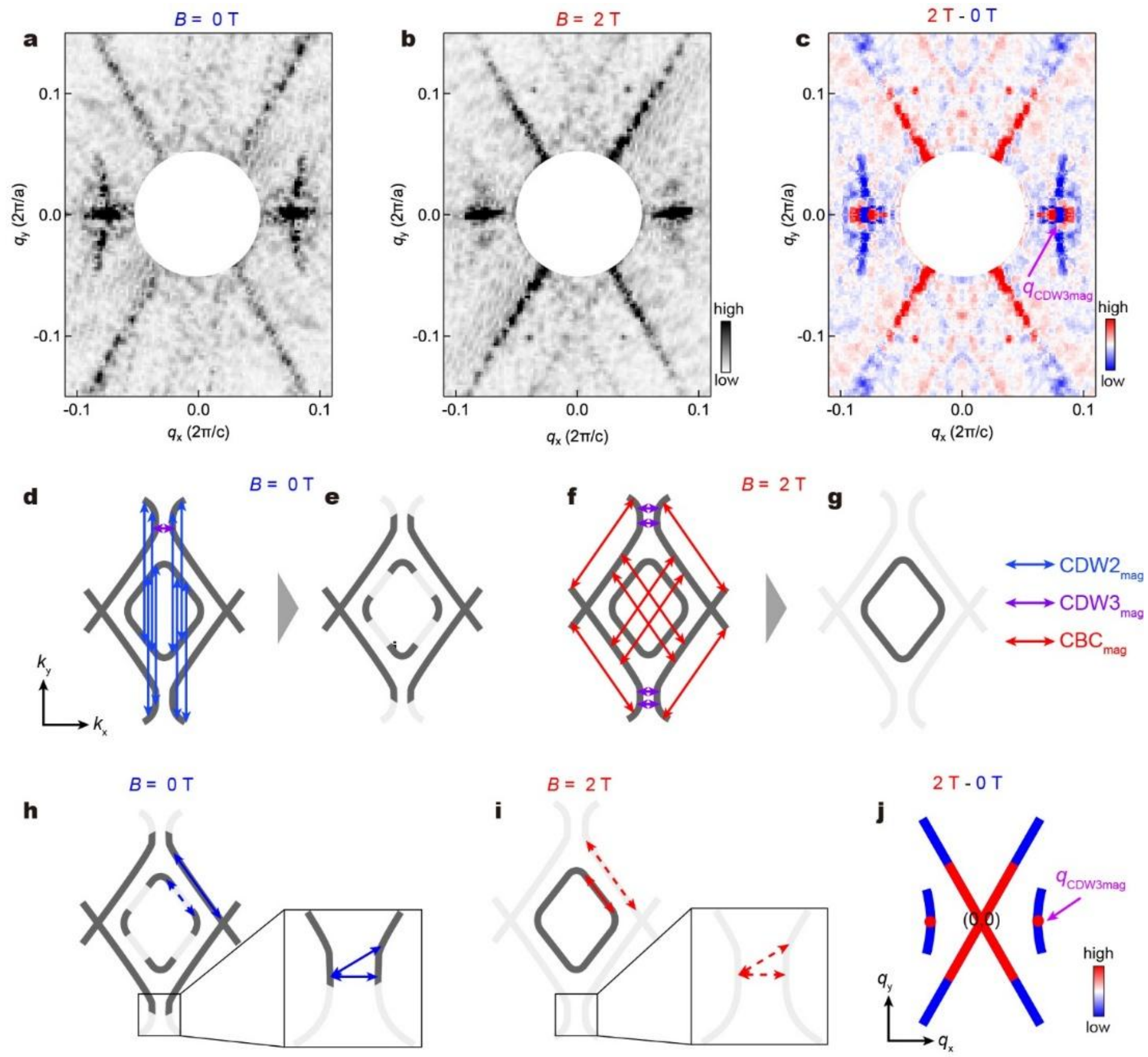

**Fig. 5. Quasiparticle interference revealing competing Fermi-surface instabilities under an in-plane magnetic field applied along the c axis.**

(a,b) Quasi-particle interference (QPI) patterns measured at (T, B) = (0.3 K, 0 T) (a), and (0.3 K, 2 T) (b). The set point is -10 mV/1 nA, and lock-in modulation is 3 mV with 961 Hz. Both were symmetrized with respect to $q_x$=0 line, assuming the quasi-tetragonal crystalline symmetry in the non-magnetic phase. Note that the color scale is identical in (a) and (b).

(c) Difference map of the QPI intensity between (0.3 K, 0 T) and (0.3 K, 2 T), showing alternating contrasts. Note that the area around q = (0,0) is masked since the major contribution of the irrelevant long wavelength signal is expected to contribute. For transparency, the non-masked data is shown in **Supplementary Note 6**. Additional QPI features in other momentum regions and their consistency with this model are discussed in **Supplementary Note 6**.

(d,e,f,g) Schematic constant-energy contour evolution owing to the gap opening by the charge orders at B=0 T (d,e) and B = 2 T (f,g), respectively. The colored arrows indicate the ordering vectors at each phase.

(h,i) Constrained scattering channels due to the gap opening at B = 0 T (f) and 2 T (g). The solid arrows indicate the possible channels, while the broken arrows indicate the suppressed channels.

(j) Expected QPI contrast by subtracting the data at B = 2 T by B = 0 T, which shows good agreement with the experiments.

**Reference**

[1] Grüner, G. Density Waves in Solids, ed Pines D (Addison–Wesley, Reading,MA) (1994).

[2] Johannes, M.D., Mazin, I.I. Fermi surface nesting and the origin of charge density waves in metals. *Phys Rev B* **77**, 165135 (2008).

[3] Y. Hong, Robertson, J.A., Kim, E-A., Kivelson, S.A. Theory of stripes in quasi-two-dimensional rare-earth tellurides. *Phys. Rev. B* **74**, 245126 (2006).

[4] Eiter, H.M., Iavagnini, M., Hackl, R., Nowadnick, E. A., Kemper, A. F., Devereaux, T. P., Chu, J-H., Analytis, J. G., Fisher, I. R., Degiorgi, L. Alternative route to charge density wave formation in multiband systems. PNAS **110** 64–69 (2013)

[5] Hamidian, M.H., Edkins, S.D., Joo, S.H. Kostin, A., Eisaki, H., Uchida, S., Lawler, M.J., Kim, E.-A., Machenzie, A.P., Fujita, K., Davis, J.C.S. Detection of a Cooper-pair density wave in $Bi_2Sr_2CaCu_2O_{8+x}$. *Nature* **532**, 343–347 (2016).

[6] Liu, X., Chong, Y.X., Sharma, R., Davis, J.C.S. Discovery of a Cooper-pair density wave state in a transition-metal dichalcogenide. *Science* **372**, 1447-1452 (2021).

[7] Aishwarya, A., May-Mann, Raghavan, A., Nie, L., Romanelli, M., Ran, S., Saha, S.R., Paglione, J., Butch, N.P., Fradkin, E., Madhavan, V. Magnetic-field-sensitive charge density waves in the superconductor $UTe_2$. *Nature* **618**, 928–933 (2023).

[8] Yin, J-X., Zhang, S.S., Li, H., Jiang, K., Chang, G., Zhang, B., Lian, B., Xiang, C., Belopolski, Ilya, Cochran, T.A., Xu, S-Y., Bian, G., Liu, K., Chang, T-R., Lin, S., Lu, Z-Y., Wang, Z., Jia, S., Wang, W., Hasan, M.Z. Giant and anisotropic many-body spin-orbit tunability in a strongly correlated Kagome magnet. *Nature* **562**, 91-95 (2018).

[9] Soumyanarayanan, A., Yee, M. M., He, Y, Wezel J. van, Rahn, D.J., Rossnagel, K., Hudson, E. W., Norman, M.R., Hoffman, J.E. Quantum phase transition from triangular to stripe charge order in $NbSe_2$. *PNAS* **110** 1623 (2013).

[10] Nagaosa, N., Tokura, Y., Topological properties and dynamics of magnetic skyrmions. *Nat. Nanotechnol*. **8**, 899 (2013).

[11] Rößler, U.K., Bogdanov, A.N., Pfleiderer, Spontaneous skyrmion ground states in magnetic metals, *Nature* **442**, 797-801 (2006).

[12] Yasui, Y., Butler, C.J., Khanh, N.D., Hayami, S., Nomoto, T., Hanaguri, T., Motome, Y., Arita, R., Arima, T., Tokura, Y., Seki, S. Imaging the coupling between itinerant electrons and localised moments in the centrosymmetric skyrmion magnet $GdRu_2Si_2$. *Nat. Commun.* **11**, 5925 (2020).

[13] Dong, Y., Arai, Y., Kuroda, K., Ochi, M., Tanaka, N., Wan, Y., Watson, M.D., Kim, T.K., Cachi, C., Hashimoto, M., Lu, D., Aoki, Y., Matsuda, T.D., Kondo, T. Fermi surface nesting driving the RKKY interaction in the

centrosymmetric skyrmion magnet $Gd_2PdSi_3$. *Phys. Rev. Lett.* **133**, 016401 (2024).

[14] Y. Dong, Y. Kinoshita, M. Ochi, Nakachi, R., Higashinaka, R., Hayami, S., Wan, Y., Arai, Y., Huh, S., Hashimoto, M., Lu, D., Tokunaga, M., Aoki, Y., Matsuda, T.D., Kondo, T. Psuedogap and Fermi arc induced by Fermi surface nesting in a centrosymmetric skyrmion magnet, *Science* **388**, 624-630 (2025).

[15] Hayami, S., Motome, Y. Noncoplanar multiple-Q spin textures by itinerant frustration: Effects of single-ion anisotropy and bond-dependent anisotropy. *Phys. Rev. B* **103**, 054422 (2021).

[16] Hayami, S., Motome, Y. Topological spin crystals by itinerant frustration. *J. Phys. Cond. Matt.* **33**, 443001 (2021).

[17] Okuma, R., Ueta, D., Kuniyoshi, S., Fujisawa, Y., Smith, B., Hsu, C. H., Inagaki, Y., Si, W., Kawae, T., Lin, H., Chuang, F. C., Masuda, T., Kobayashi, R., Okada, Y., Fermionic order by disorder in a van der Waals antiferromagnet. *Scientific Reports* **10**, 15311 (2020)

[18] Okuma, R., Ritter, C., Nilsen, G.J., Okada, Y. Magnetic frustration in a van der Waals metal CeSiI. *Phys. Rev. Materials* **5**, L121401 (2021).

[19] Posey, V.A., Turkel, S., Rezaee, M., Devarakonda, A., Kundu, A.K., Ong, C.S., Thinel, M., Chica, D.G., Vitalone, R.A., Jing, R., Xu, S., Needell, D.R., Meirzadeh, E., Feuer, M.L., Jindal, A., Cui, X., Valla, T., Thunström, P., Yilmaz, T., Vescovo, E., Graf, D., Zhu, X., Scheie, A., May, A.F., Eriksson, O., Basov, D.N., Dean, C.R., Rubio, A., Kim, P., Ziebel, M.E., Millis, A.J., Pasupathy, A.N., Roy, X. Two-dimensional heavy fermions in the van der Waals metal CeSiI. *Nature* **625**, 483–488 (2024).

[20] Okuma, R., Yamagami, K., Fujisawa, Y., Hsu, C.H., Obata, Y., Tomoda, N., Dronova, M., Kuroda, K., Ishikawa, H., Kawaguchi, K., Aido, K., Kindo, K., Chan, Y.H., Lin, H., Ihara, Y., Kondo, T. Okada, Y. Emergent topological magnetism in Hund's excitonic insulator. *arXiv* 2405. 16781 (2024).

[21] Imada, M., Fujisamori, A., Tokura, Y. Metal-insulator transitions. *Rev. Mod. Phys*. **70**, 1039 (1998).

[22] Geim, A. K. & Grigorieva, I. V. Van der Waals heterostructures. *Nature* **499**, 419–425 (2013).

[23] Yumigeta, K., Qin, Y., Li, H., Blei, M., Attarde, Y., kopas, C., Tongay, S. Advances in Rare-Earth Tritelluride Quantum Materials: Structure, Properties, and Synthesis. *Adv. Sci.* **8**, 2004762 (2021).

[24] Chen, D., Zhang, S., Yang, H.X., Li, J.Q., and Chen, G.F. Magnetic and transport properties of a layered compound $Ce_2Te_5$. *J. Phys.: Condens. Matter* 29 265803 (2017).

[25] Wang, P.P., Long, Y-J., Zhao, L.-X., Chen, D., Xue, M-Q., Chen, G.F. Anisotropic transport and magnetic properties of charge-density-wave materials $R$SeTe$_2$ ($R$ = La, Ce, Pr, Nd). *Chin. Phys. Lett.* **32**, 087101 (2015).

[26] Yumigeta, K., Kopaczek, J., Attarde, Y., Sayyad, M.Y., Blei, M., Moosavy, S.T.R.J., Yarra, A., Ruddick, H., Povilus, B., Banerjee, R., Ou, Y., Tongay, S. Alloying Effect of Rare-Earth Tritellurides on the Charge Density Wave and Magnetic Properties. *Appl. Phys. Rev.* **11**, 011407 (2024)

[27] Lei, S., Lin, J., Jia, Y., Gray, M., Topp, A., Farahi, G., Klemenz, S., Gao, T., Rodolakis, F., McChesney, J. L., Ast, C.R., Yazdani, A., Burch, K. S., Wu, S., Ong, N. P., Schoop, L. M. High mobility in a van der Waals

layered antiferromagnetic metal. *Sci. Adv.* **6**, eaay6407 (2020).

[28] Watanabe, M., Nakamura, R., Lee, S., Asano, T., Ibe, T., Tokuda, M., Taniguchi, H., ueta, D., Okada, Y., Niimi, Y. Shubnikov-de Haas oscillation and possible modification of effective mass in CeTe3 thin films. *AIP Adv.* **11**(1) 015005 (2021).

[29] Higashihara, T., Asama, R., Nakamura, R., Watanabe, M., Tomoda, N., Hasiweder, T.J., Fujisawa, Y., Okada, Y., Iwasaki, T., Watanabe, K., Taniguchi, T., Jiang, N., Niimi, Y. Magnetotransport properties in van der Waals $R$Te3 ($R$ =La, Ce, Tb). *Phys. Rev. B* **109**, 134404 (2024)

[30] Straquadine, J.A.W., Ikeda, M.S., Fisher, I.R. Evidence for Realignment of the Charge Density Wave State in ErTe3 and TmTe3 under Uniaxial Stress via Elastocaloric and Elastoresistivity Measurements. *Phys. Rev. X* **12**, 021046 (2022).

[31] Singh, A.G., Bachmann, M.D., Sanchez, J.J., pandey, A., Kapitulnik, A., Kim, J.W., Ryan, P.J., Kivelson, S.A., Fisher, I.R. Emergent tetragonality in a fundamentally orthorhombic material. *Sci. Adv.* **10**, eadk3321 (2024).

[32] Gallo–Frantz, A., Jacques, V.L.R., Sinchenko, A.A., Ghoneim, D., Ortega, L., Godard, P., Renault, P.-O., Hadj-Azzem, A., Lorenzo, J.E., Monceau, P., Thiaudière, D., Grigoriev, P.D., Bellec, Bolloc'h, E., D. Le, Charge density waves tuned by biaxial tensile stress. *Nat. Commun*. **15**, 3667 (2024).

[33] Kogar, A., Zong, A., Dolgirev, P.E., Shen, X., Straquadine, J., Bie, Y.-Q., Wang, X., Rohwer, T., Tung, I-C., Yang, Y., Li, R., Yang, J., Weathersby, S., Park, S., Kozina, M.E., Sie, E.J., Wen, H., Jarillo-Herrero, P., Fisher, I.R., Gedik, N., Light-induced charge density wave in $LaTe_3$. *Nat. Phys.* **16** 159 (2020).

[34] Sinchenko, A.A., Lejay, P., Monceau, Sliding charge-density wave in two-dimensional rare-earth tellurides. *Phys. Rev. B* **85**, 241104(R) (2012).

[35] Hamlin, J.J., Zocco, D.A. Sayles, T.A., Maple, M.B., Chu, J., Fisher, I.R. Pressure-induced superconducting phase in the charge-density-wave compound Terbium tritelluride. *Phys. Rev. Lett.* **102**, 177002 (2009).

[36] Zacco, D.A., Hamlin, J.J., Grube, K., Chu, J.H., Kuo, H.H., Fisher, I.R., Maple, M.B., Pressure dependence of the charge-density-wave and superconducting states in $GdTe_3$, $TbTe_3$, and $DyTe_3$. *Phys. Rev. B* **91**, 205114 (2015).

[37] Li, J., Feng, J., Wang, D., Peng, S., Li, M., Wang, H., Xu, Y., Xhao, T., Zhao, B., Jiang, S., Li, X., Lin, C., Li, Y. Pressure-induced structural evolution with suppression of the charge density wave state and dimensional crossover in $CeTe_3$, *Phys. Rev. B* **109**, 094119 (2024).

[38] Sarkar, S., Bhattacharya, J., Sadhukhan, P., Curcio, D., Dutt, R., Singh, V.K., Bianchi, M., Pariari, A., Roy, S., Mandal, P., Das, T., Hofmann, P., Chakrabarti, A., Barman, S.R., Charge density wave induced nodal lines in $LaTe_3$. *Nat. Commun.* **14**, 3628 (2023).

[39] Wang, Y., Petrides, I., McNamara, G., Hosen, M.M., Lei, S., Wu, Y.-C., Hart, J.L., Lv, H., Yan, J., Xiao, D., Cha, J.J., Narang, P., Schoop, L. M., Burch, K.S. Axial Higgs mode detected by quantum pathway interference in $RTe_3$. *Nature* **606**, 896–901 (2022).

[40] Wulferding, D., park, J., Tohyama, T., Park, S.R., Kim, C. Magnetic field control over the axial character of Higgs modes in charge-density wave compounds. *Nat. Commun.* **16**, 114 (2025).

[41] Nakamura, T., Fujisawa, Y., Smith, B.R.M., Tomoda, N., Hasiweder, T.J., Okada, Y. Revealing pronounced electron-hole Fermi pockets in the charge density wave semimetal $LaTe_3$. *Phys, Rev. B* **110** 235415 (2024).

[42] Eiter, H.M., Lavagnini, M., Hackl, R., Nowadnick, E.A., Kemper, A.F., Devereaux, T.P., Chu, J.H., Analytis, J.G., Fisher, I.R., Degiorgi, L. Alternative route to charge density wave formation in multiband systems. *PNAS* **110**, 64-69 (2013).

[43] Chillal, S. Schierle, E., Weschke, E., Yokaichiya, F., Hoffmann, J.-U., Volkova, O.S., Vasiliev, A.N., Sinchenko, A.N., Lejay, P., Hadj-Azzem, A., Monceau, P., Lake, B. Strongly coupled charge, orbital, and spin order in $TbTe_3$. *Phys. Rev. B* **102**, 241110(R) (2020).

[44] Akatsuka, S., Esser, S., Okumura, S., Yambe, Ryota, Yamada, Y., Hirschmann, M.M., Aji, S., White, J.S., Gao, S., Onuki, Y., Arima, T., Nakajima, T., Hirschberger, M. Non-coplanar helimagnetism in the layered van-der-Waals metal $DyTe_3$. *Nat. Commun.* **15**, 4291 (2024).

[45] Raghavan, A., Romanelli, M., May-Mann, J., Aishwarya, A., Aggarwal, L., Singh, A.G., Bachmann, M.D., Schoop, L.M., Fradkin, E., Fisher, I.R., Madhavan, V. Atomic-scale visualization of a cascade of magnetic orders in the layered antiferromagnet $GdTe_3$, *npj Quantum Materials* **9**, 47 (2024).

[46] Aoyama, K. RKKY interaction in the presence of a charge density wave order, *Phys. Rev. B* **111**, L100404 (2025)

[47] Ru, N., Fisher, I.R. Thermodynamic and transport properties of $YTe_3$, $LaTe_3$, and $CeTe_3$, *Phys. Rev. B* **73**, 033101 (2006).

[48] Deguchi, K., Okada, T., Chen, G.F., Ban, S., Aso, N., Sato, N.K. Magnetic order of rare-earth tritelluride $CeTe_3$ at low temperature. *J. Phys.: Conf. Ser.* **150** 042023 (2009).

[49] Ueta, D., Kobayashi, R., Sawada, H., Iwata, Y., Yano, S., Kuniyoshi, S., Fujisawa, Y., Masuda, T., Okada, Y., Itoh, S. Anomalous magnetic moment direction under magnetic anisotropy originated from crystalline electric field in van der Waals compounds $CeTe_3$ and $CeTe_2Se$. *J. Phys. Soc. Jpn*. 91, 094706 (2022).

[50] Okuma, R., Fujisawa, Y., Maekawa, N., Nakao, A., Ishikawa, Y., Kobayashi, R., Okada, Y., and Ueta, D., Quantum fluctuation driven double-q magnetic order in the van der Waals metal $CeTe_3$, submitted to *Phys. Rev. B.*

[51] Brouet, V, Yang, W.L., Zhou, X.J., Hussain, Z., Ru, N., Shin, K.Y., Fisher, I.R., Shen, Z.X. Fermi surface reconstruction in the CDW state of $CeTe_3$ observed by photoemission. *Phys. Rev. Lett.* **93**, 126405 (2004).

[52] Brouet, V., Yang, W.L., Zhou, X.J., Hussain, Z., Moore, R.G., He, R., Lu, D.H., Shen, Z.X., Laverock, J., Dugdale, S.B., Ru, N., Fisher, I.R. Angle-resolved photoemission study of the evolution of band structure and charge density wave properties in $RTe_3$ (R=Y, La, Ce, Sm, Gd, Tb, and Dy). *Phys. Rev. B* **77**, 235104 (2008).

[53] Smith, B.R.M., Fujisawa, Y., Wu, P., Nakamura, T., Tomoda, N., Kuniyoshi, S., Ueta, D., Kobayashi, Okuma, R., Arai, K., Kuroda, K., Hsu, C-H., Chang, G., Huang, C-Y., Lin, H., Wang, Z.Y., Kondo, T., Okada, Y.

Uncovering hidden Fermi surface instabilities through visualizing unconventional quasiparticle interference in $CeTe_3$. *Phys. Rev. Mat.* **8**, 104004 (2024).

[54] Zang, H, et al., Kondo-coupled van der Waals antiferromagnet with high-mobility quasiparticles, Newton 2, 100320 (2026).

[55] Trontl, V.M., et al., Interplay of Kondo Physics with incommensurate Charge Density Wave in $CeTe_3$, arXiv 2502. 04814v1 (2025).

[56] Hayami, S. Anisotropic skyrmion crystals on a centrosymmetric square lattice under an in-plane magnetic field. J. *Mag. Mag. Mat*. 604, 172293 (2024).

[57] Jérome, D., Rice, T.M., Kohn, W. Excitonic insulator. *Phys. Rev.* **158** 462 (1967).

[58] Keldysh, L. V., and Kopaev, Y.V. Possible instability of the semimetallic state toward Coulomb interaction. *Selected Papers of Leonid V Keldysh*. 41-46 (2024).

[59] Xie, Y.M., Nagaosa, N., Phase shifts, band geometry, and responses in triple-Q charge and spin density waves, *Phys. Rev. B* **110**, L241108 (2024).

[60] E. Andrei et al., *The marvels of moiré materials,* Nature Reviews Materials 6, 201 (2021).

[61] D. M. Kennieth et al., *Moiré heterostructures as a condensed-matter quantum simulator*
Nature Physics **17**,155 (2021).

[62] Ru, N., Condron, C.L., Maggulis, G.Y., Shin, K.Y., laverock, J., Dugdale, S.B., Toney, M.F., and Fisher, I.R., Effect of chemical pressure on the charge density wave transition in rare-earth tritellurides $R\mathrm{Te}_3$, *Phys. Rev. B* **77**, 035114 (2008).

[63] Hu, B.F., Cheng, B., Yuan, R.H., Dong, T., and Wang, N.L., Coexistence and competition of multiple charge-density-wave orders in rare-earth tritellurides, *Phys. Rev. B* **90**, 085105 (2014).

[64] Banerjee, A., Feng, Y., Silevitch, D.M., Wang, J., Lang, J.C., Kuo, H.-H, Fisher, I.R., Rosenbaum, T.F., Charge transfer and multiple density waves in the rare earth tellurides, Phys. Rev. B 87, 155131 (2013).

[65] Ru, N., Chu, J.-H., Fisher, I.R., Magnetic properties of the charge density wave compounds $R\mathrm{Te}_3$ ($R$=Y, La, Ce, Pr, Nd, Sm, Gd, Tb, Dy, Ho, Er, and Tm), *Phys. Rev. B* **78**, 012410 (2008).